\documentclass[%
reprint,
superscriptaddress,
amsmath,amssymb,
aps,
floatfix,
]{revtex4-2}

\usepackage{graphicx}
\usepackage{dcolumn}
\usepackage{bm,bbm}
\usepackage[dvipsnames]{xcolor}
\usepackage[normalem]{ulem}
\usepackage{enumitem}
\usepackage{hyperref}

\newcommand{\1}{\mathbbm{1}}

\newcommand{\be}{\begin{equation}}
\newcommand{\ee}{\end{equation}}
\newcommand{\bea}{\begin{eqnarray}}
\newcommand{\eea}{\end{eqnarray}}
\newcommand{\bes}{\begin{equation*}}
\newcommand{\ees}{\end{equation*}}
\newcommand{\beas}{\begin{eqnarray*}}
\newcommand{\eeas}{\end{eqnarray*}}
\usepackage{color}

\definecolor{james}{rgb}{1,.6,0}

\begin{document}

\title{A highly efficient tensor network algorithm for multi-asset Fourier options pricing}

\author{Michael Kastroyano}
\email{kastorm@amazon.com}
\affiliation{Amazon Quantum Solutions Lab, Seattle, Washington 98170, USA}
\affiliation{AWS Intelligent and Advanced Compute Technologies, Professional Services, Seattle, Washington 98170, USA}
\affiliation{AWS Center for Quantum Computing, Pasadena, CA 91125, USA}
 
\author{Nicola Pancotti}
\affiliation{AWS Center for Quantum Computing, Pasadena, CA 91125, USA}

\date{\today}

\begin{abstract}

  Risk assessment and in particular derivatives pricing is one of the
  core areas in computational finance and accounts for a sizeable
  fraction of the global computing resources of the financial industry.
  We outline a quantum-inspired algorithm for multi-asset options
  pricing.
     The algorithm is based on tensor networks, which have
  allowed for major conceptual and numerical breakthroughs in quantum
  many body physics and quantum computation. In the proof-of-concept
  example explored, the tensor network approach yields several orders of
  magnitude speedup over vanilla Monte Carlo simulations.  We take this
  as good evidence that the use of tensor network methods holds great
  promise for alleviating the computation burden of risk evaluation in
  the financial and other industries, thus potentially lowering the
  carbon footprint these simulations incur today.

\end{abstract}

\maketitle

\section{Introduction}

Options pricing remains an active field of research and development in
financial engineering. In its simplest form, an option is a contract
which allows the buyer to purchase some specified underlying financial
asset, such as stock, bonds, etc.~at a given {\em strike} price at a
specified expiration date. The central task of options pricing is to
evaluate what the fair price is for such a contract, such that
arbitrage---the possibility to make a profit at no risk---is impossible.
A great number of variants to this simple setting exist, including
functions of several assets, multiple expiration dates, and time
dependent returns \cite{hullBook}. Derivative products are essential for
investors seeking to expand their investment and hedging strategies, and
have constituted an especially fluid market for decades. Given the large
volume and diversity of derivatives products being traded on the
financial markets, understanding their properties is an essential
objective of financial engineering.

Only the very simplest options pricing scenarios can be evaluated
analytically. Therefore, in general, one has to resort to numerical
simulations.  These typically break up into two general approaches:
Either approximate the solution via Monte Carlo or Fourier methods,
building upon the general Feynman-Kac formulation of the solution, or
map the stochastic differential equation (SDE) describing the dynamics
to a partial differential equation (PDE), and subsequently solve the
continuous PDE with appropriate boundary conditions and a discretization
scheme. In this work we do not consider the PDE-based approach.

Monte Carlo methods are generally simple to implement and are very
versatile, but have a rather slow sample convergence rate that is
proportional to the inverse square root of measurements $n$, i.e.~of
${\mathcal O}(1/\sqrt{n})$.  Quasi-Monte Carlo methods can in certain
cases improve the scaling to close to ${\mathcal O}(1/n)$ \cite{quasiMC,
mlmc}, but their accuracy and reliability is on a case by case basis.
Monte Carlo methods also have the significant advantage that they
typically do not suffer from the curse of dimensionality---that is, the
effort scales exponentially in the space dimension---and are thus the
only numerical method available for multi-asset pricing scenarios.
Furthermore, Monte Carlo methods also struggle with complex
path-dependent options where the computations can no longer be
parallelized easily.

The main focus in this article is on the Fourier method of options
pricing. This numerical approach is typically many orders of magnitude
faster than any other on single assets, when applicable. For this
reason, it is sometimes referred to as an {\em exact method}. However,
it suffers severely from the curse of dimensionality. See
Ref.~\cite{benchop} for standard comparisons of the different numerical
approaches to simple options of a single asset.

A number of proposals have recently appeared \cite{Q1,Q2,Q3,Q4,Q5}
suggesting a quantum advantage for the options pricing problem. Most of
them are based on the work by Montanaro \cite{montanaro} showing that
Monte Carlo sampling can be performed on a quantum computer with an
${\mathcal O}(1/n)$ convergence rate. However, the algorithm requires a
fault-tolerant quantum computer because it is based on quantum phase
estimation (QPE), which requires a circuit depth that is well beyond the
reach of currently-available noisy intermediate-scale quantum (NISQ)
devices.  Furthermore, it is unclear whether the QPE approach will ever
be competitive given the modest polynomial scaling advantage, the large
overhead of fault-tolerant quantum computing and the competing classical
quasi-Monte Carlo methods. Chakrabarti {\em et.~al} \cite{Q3} argue that
for exotic options there will be a threshold where QPE provides an
advantage which is of the same order of magnitude as that argued for in
quantum chemistry. A different approach proposes to solve the PDE
associated with an option \cite{Q4}, yet, in view of a careful analysis
of quantum algorithms for differential equations
\cite{childs,montanaroFEM}, falsely claim an exponential speedup.

In this paper we consider an approach to the multi-asset options pricing
problem that is loosely inspired by quantum computing. We start from the
Fourier formulation of the multi-asset problem and formulate it as the
inner product of two complex vectors: one representing the
characteristic function of the stochastic dynamics and one representing
the Fourier transform of the payoff. The two vectors are
approximately built as nonnormalized matrix product states (MPS) via the
TT-cross \cite{TTcross1} algorithm. The inner product can then be
evaluated efficiently and thus multi-asset options pricing accelerated
accordingly.

On a correlated multi-asset example, we find that our algorithm
accurately matches the exact results up to $4$ assets with dramatically
reduced computational time. Beyond that it matches Monte Carlo-based
results and provides a significant speedup for up to $15$ assets.  Our
approach can be applied to many practical options pricing scenarios, and
therefore holds great promise for practical computational speedup.

The paper is structured as follows. Section \ref{sec:eco} introduces the
European call option, followed by an outline of Monte Carlo and
Fourier-based options pricing in Sec.~\ref{sec:mcf}. Section
\ref{sec:TN} then presents the quantum formulation of the problem, as
well as a solution using tensor networks. Numerical experiments are
shown in Sec.~\ref{sec:num}, followed by concluding remarks.

\section{The European call option}
\label{sec:eco}

The European call option is a contract established between a buyer and a
seller relating to an underlying asset (bond, stock, interest rate,
currency, etc.). The contract allows the buyer to purchase the underlying
asset at a specific strike price $K$ at the termination date $T$. The
buyer has no obligation to purchase the underlying asset. Therefore,
given an asset price $S_0$ at time $t=0$, a strike price $K$ and a
termination date $T$, the value of the option at time $T$ is
\be
v(S_T,K)= \max\{S_T-K,0\},
\ee
for the specific (random) trajectory of the asset starting at price
$S_0$ and ending at price $S_T$. The main challenge in options pricing
is to determine the value of the option at time $t\leq T$. The
Feynman-Kac formula provides a mathematically precise answer to this
problem, namely

\be
V(t,K|S_0)= e^{-r(T-t)}\mathbb{E}[v(S_T,K)|S_0],\label{Feynmann}
\ee
where $e^{-r(T-t)}$ accounts for a constant discount $r$ with a
risk-free investment such as a bond. The expectation is over all
stochastic trajectories of the underlying asset starting at $S_0$. The
discounted price, Eq.~(\ref{Feynmann}) is a Martingale measure under the
no-arbitrage assumption \cite{hullBook}.

We start with the simplest model, namely the Black-Scholes model with
one underlying asset. The model is defined either in SDE or PDE setting.
Let $S_t$ be the value of an underlying asset at time $t\geq 0$. Under
geometric Brownian motion, the time evolution is described by the SDE
\be
dS_t = r S_t dt + \sigma S_t dW_t, \label{eq:sde1}
\ee
where $r$ is a constant rate of return, and $\sigma$ a variance called
the volatility. Black and Scholes showed \cite{blackscholes} that the
dynamics of the European put option under geometric Brownian motions
can be cast as the PDE of the form
\be
\frac{\partial V(t,S)}{\partial t}+\frac{1}{2}\sigma^2 S^2 
\frac{\partial^2 V(t,S)}{\partial S^2}+rS 
\frac{\partial V(t,S)}{\partial S}=0,
\ee
where $V(t,S)$ describes the price of the asset at time $t>0$, subject
to the termination condition
\be 
V(T,S_T)= \max\{S_T-K,0\}.
\ee
In the PDE setting one wants to solve the inverse time problem, i.e.,
given a boundary condition at time $t=T$, find the solution at time
$t=0$. In this very special case, there exists an exact solution (known
as the the Black Scholes solution) given by
\be 
V(t,S_0) = S_0 \Phi(d_1) - K e^{-r(T-t)}\Phi(d_2)\label{eq:BSexact},
\ee
where
\be
 \Phi(x) = \frac{1}{\sqrt{2\pi}}\int_{-\infty}^x e^{-s^2/2}ds
\ee 
and
\be
 d_{1/2}= \frac{\ln(S_0/K)+(r\pm \sigma^2/2)(T-t)}{\sigma \sqrt{T-t}}.
\ee
In the early days of electronic options trading, most options were
priced using this formula because it was accurate and simple to compute.
However, following the stock market crash of 1989, practitioners
realised that the Black-Scholes model did not account for the
``volatility smile,'' rather assuming that the volatility is constant
and independent of the strike price. More sophisticated models are used
today that better account for the heavy tailed distribution of financial
assets, such as, for example, the SABR model \cite{SABR}.  However, the
Black-Scholes model remains the conceptual building block behind much of
options pricing. Unfortunately, the more sophisticated models and payoff
functions do not allow for analytic solutions and one must resort to
numerical estimations.

\subsection{Multi-asset extension}

We consider a multi-asset extension of the European put option under
geometric Brownian motion. Let $\vec{S_t}=\{ S_t^1, ..., S_t^d\}$ denote
the prices of $d$ underlying assets. The geometric Brownian motion can
similarly be extended to the multi-asset scenario via
\be
dS^j_t = r S^j_t dt + \sigma_j S^j_t dW^j_t, 
\ee
where the stochastic terms $\{dW_t^j\}$ have Gaussian correlations
described by the correlation matrix 
\be
\mathbb{E}[dW^j_tdW^j_t]= \Sigma_{ij} dt.
\ee 
We assume that the correlation matrix is positive definite throughout.

We consider a commonly-used multi-asset option, the min option, which
returns the minimum of the underlying assets for a fixed strike price $K$:
\be 
v_{\min}(S_T,K) = \max\{\min\{S_T^1,...,S_T^d\}-K, 0\}.\label{eq:minpayoff}
\ee
There are many other possible extensions to multiple assets. However,
this is the simplest one that captures correlations both in the
stochastic dynamics, as well as in the multi-asset option. In the
correlated multi-asset case there no longer exists an analytic solution,
so numerical simulations are needed.  In practice, Monte Carlo methods
are often used, because they do not suffer from the curse of
dimensionality.

\section{Monte Carlo and Fourier simulations}
\label{sec:mcf}

\subsection{Monte Carlo options pricing}\label{sec:MC}

In this section we succinctly describe the Monte Carlo and Fourier
approaches to numerical options pricing. We start with the Feynman-Kac
formula, and express the expectation as an integral over the conditional
probability $p(S_T|S_0)$ that the value of the asset at time $T$ is
$S_T$ given that its value at time $t=0$ is $S_0$:
\bea 
V(t,K|S_0)&=&e^{-r(T-t)}\int_0^\infty v(S_T,K) p(S_T|S_0)dS_T\\
&=&e^{-r(T-t)}\int_{-\infty}^\infty v(e^x,K) p(x|x_0)dx, 
\label{eq:mc}
\eea
where in the second line we changed variables to $x=\ln(S_T)$ and
$x_0=\ln(S_0)$. The change of variables reduces geometric Brownian
motion to regular Brownian motion, where $p(x|x_0)$ is the normal
distribution with mean at $\mu = s_0/T+r-\sigma^2/2$ and standard
deviation $\sigma$. The Monte Carlo approach to evaluate the integral in
Eq.~(\ref{eq:mc}) is by sampling geometric Brownian motion trajectories
according to Eq.~(\ref{eq:sde1}), and averaging over the functional
outcomes $v(e^x,K)$:
\begin{itemize}[leftmargin=2.5em]

\item[(a)] Simulate a {\em trajectory}  ${\bf S}=\{S_0,S_{\delta t},
\cdots, S_T\}$, where $M=T\delta t$ is the number time points, by the
discretization  of the SDE in Eq.~(\ref{eq:sde1}).

\item[(b)] Generate $n$ trajectories (samples) $\{{\bf
S}^{(j)}\}_{j=1}^N$, and evaluate the function $v(S_T,K)$ for each
trajectory.

\item[(c)] Approximate \be \int_{-\infty}^\infty v(e^x,K) p(x|x_0)dx \approx \frac{1}{N}\sum_{j=1}^N v(S^{(j)}_T,K)\nonumber \ee

\end{itemize}
While extremely natural and general for path independent options, the
Monte Carlo approach converges rather slowly proportional to
$O(1/\sqrt{n})$, where $n$ is the number of samples, making it
computationally costly to obtain high-precision answers, especially when
multiple strike prices are desired.  Monte Carlo methods also struggle
with time dependent options, where independent trajectories can no
longer be used throughout the lifetime of the option \cite{barrier}.

\subsection{Fourier options pricing}

In Fourier options pricing, the probability density $p(x|x_0)$ is
constructed explicitly from the characteristic function $\varphi_T$ of
the dynamics, via the relation
\be 
\varphi_T(u)=\mathbb{E}[e^{i u X_T}|X_0]=\int_{\mathbb{R}} e^{i u x} p(x|x_0) dx,
\label{eq:cf} 
\ee 
where $X_t$ is the random variable at time $t$ of the stochastic process
described by Eq.~(\ref{eq:sde1}).

The reason for working with the characteristic function is twofold.
First, the conditional probability $p(x|x_0)$ is only known analytically
in very few special cases, while analytic expressions for the
characteristic function are more widely known (e.g., Black-Scholes-Merton
\cite{lewis}, Heston \cite{heston}) or can be approximated (e.g., SABR
\cite{SABR,van2018cos}). Second, the function is more stable in the
Fourier domain for small time steps due to the uncertainty principle.

We can substitute the inverse of Eq.~(\ref{eq:cf}) in Eq.~(\ref{eq:mc})
to obtain:
\be 
V(t,K|S_0) = e^{-r(T-t)}\int_{\mathbb{R}^2} e^{-iux}v(e^x,K)\varphi_T(u) dxdu.
\label{eq:naive}
\ee
This expression can be approximated numerically, but it suffers a rather
large discretization error because $v(e^x,K)$ is not square integrable
in $\mathbb{R}$. To remedy this, Carr and Madan \cite{carrmadan} have
proposed a method of stabilising the  nonintegrability by shifting the
Fourier integration into the complex plane. We outline a slightly
different approach due to Lewis \cite{lewis}, which generalizes more
easily to the types of multi-dimensional payoffs that we consider.

The Lewis approach is based on taking the Fourier transform of the
payoff function. Consider the case of the European call option $v_{\rm
call} (z)$. Its Fourier transform is given by 
\bea 
\hat{v}_{\rm call} (z) &=& \int_{-\infty}^\infty e^{izx}\min\{e^x-K,0\} dx\\
&=&\left. \left(\frac{e^{(iz+1)x}}{iz+1}- K \frac{e^{izx}}{iz}\right) 
\right|_{\ln(K)}^\infty= \frac{-K^{iz +1}}{z (z-i)},
\eea
where $z$ is complex, and the last equation only holds in the strip
${\rm Im}[z]>1$, because the Fourier transform of the payoff function
has a branch cut in the complex plane. The inverse transform can then be
expressed as
\be 
v_{\rm call}(x) = \frac{1}{2\pi}
	\int_{ -\infty+\alpha}^{\infty+i\alpha}e^{-izx}\hat{v}_{\rm call}(z) dz,
\ee
where $\alpha>1$ guarantees that the Fourier transform of the payoff is
well defined. Clearly, different payoffs will have different domains of
integrability. For a comprehensive analysis, see
Refs.~\cite{schmelzle,eberlein}. We can then express the options price
as:
\bea 
V(t,K|S_0) &=& e^{-r (T-t)} \mathbb{E}[v_{\rm call}(X_T)|X_0]\nonumber\\
 &=&  \frac{e^{-r (T-t)}}{2\pi} \mathbb{E}[\int_{i\alpha -\infty}^{i \alpha+\infty}e^{-izX_T}\hat{v}_{\rm call}(z) dz|X_0]\nonumber\\
 &=& \frac{e^{-r (T-t)}}{2\pi} \int_{i\alpha -\infty}^{i \alpha+\infty}\mathbb{E}[e^{-izX_T}|X_0]\hat{v}_{\rm call}(z) dz\nonumber\\
 &=& \frac{e^{-r (T-t)}}{2\pi} \int_{i\alpha -\infty}^{i \alpha+\infty}\varphi_T(-z)\hat{v}_{\rm call}(z) dz\label{eq:maincont}
\eea

\subsection{The multi-asset case}

The Fourier method can easily be extended to the multi asset case
provided the multi-asset characteristic function is known and the
Fourier transform of the payoff can be evaluated explicitly.  The
characteristic function of multi-asset geometric Brownian motion is
given by:
\be 
\varphi_T(\vec{z})=\exp\left[ i\sum_j^d z_j \mu_j -
\frac{1}{2}\sum_j^d\sum_k^d \sigma_j\sigma_kz_j z_k \Sigma_{jk})\right],
\ee
where
\be
\mu_j = x^j_0 +r_jT -\frac{1}{2} \sigma_j^2 \Sigma_{jj}T,
\ee
with $j=1,...,d$.  The Fourier transform of the min option is
\cite{eberlein}:
\be 
\hat{v}_{\rm min}(\vec{z})=
\frac{K^{1+i\sum_{j=1}^dz_j}}{(-1)^d(1+i\sum_{j=1}^dz_j)\prod_{j=1}^d z_j},
\ee
and is subject to the conditions: ${\rm Im}[z_j]>0$ and $\sum_{j=1}^d
{\rm Im}[z_j]>1$. Note that, $\hat{v}_{\min}$ reduces to the single
asset formula when $d=1$.  We can express the price of the multi-asset min
option as
\be  
V(t,K|S_0)=\frac{e^{-r (T-t)}}{2\pi}
 \int_{\mathbb{R}^d+i\vec{\alpha} }\varphi_T(-\vec{z})\hat{v}_{\rm min}(\vec{z}) d\vec{z}.\label{eq:main}
\ee

\subsection{Discretization}

We now discuss the numerical evaluation of the single asset formula in
Eq.~(\ref{eq:maincont}). The multi-asset extension is straightforward.
We first need to truncate the integral at a finite value. The
probability distribution $p(x|x_0)$ is Gaussian, so it can be well
approximated on a finite square of size $[-a \sigma,a\sigma]$ around the
mean, where the approximation improves exponentially in $a$. Typically
$a=5$ yields good numerical precision. The discretization in real and
Fourier space should satisfy the uncertainty relation: $dx \cdot dk
=2\pi/N$. Given that $dx\propto 1/N$, we expect the Fourier
increment to be constant. $dk$ is left as a free variable in our
numerical experiments, and is typically taken to be of order $1$. Its
value can  heavily affect the accuracy of the result. The expression to
evaluate is then
\be 
V(t,K|S_0)\!=\!\frac{e^{-r(T-t)}}{2\pi}
\!\!\!\!\!\!
\sum_{j=-N/2}^{N/2}\!\!\!\varphi_T(-\eta j-i\alpha)\hat{v}_{\rm min}(\eta j+i\alpha)\eta,\label{eq:discr}
\ee
where $\eta$ is identified with $dk$. We pause here to
make an observation about the shift to the imaginary plane. One might
naively expect that Eq.~(\ref{eq:discr}) is equivalent to the
discretization of Eq.~(\ref{eq:mc}). Indeed, one could perform the shift
to the complex plane as well as the discrete Fourier transform. However,
the catch is that the discrete analogue of the delta function
$(1/2\pi)\sum_j \exp[ij\eta(x-y)]$ is non-zero away from $x=y$, and
gives rise to an error. In this way there is a very subtle numerical
error that creeps up in the original expression Eq.~(\ref{eq:mc}), that
has its roots in complex analysis.

Equation (\ref{eq:discr}) yields exquisite precision already for small
values of $N$ (see Sec.~\ref{sec:num}). For that reason, it is often
referred to as a quasi-exact method in the options pricing literature.
Numerically simulating the multi-asset extension of Fourier pricing is
straightforward by replacing the single sum by a nested sum. However,
the evaluation of the nested sum becomes computationally prohibitive
beyond a few assets. Indeed, building the vectors $\vec{\varphi}_T$ and
$\hat{\vec{v}}_{\rm min}$ also becomes prohibitive beyond a few assets,
scaling exponentially in the number of assets.

\section{Quantum formulation and tensor network solution}
\label{sec:TN}

We propose a quantum-inspired solution based on tensor networks which
reduces the computational cost of multi-asset Fourier options pricing
from ${\mathcal O}(N^d)$ to ${\mathcal O}(d N)$. As shown in
Sec.\ref{sec:num}, our method typically has the same asymptotic scaling
as Monte Carlo, but with a prefactor several orders of magnitude
smaller.  The main idea is to express Eq.~(\ref{eq:main}) or
Eq.~(\ref{eq:mc}) as the inner product of two (nonnormalized) quantum
states in matrix product form. In the case of Eq.~(\ref{eq:main}) we
obtain:
\be 
V(t,K|S_0)=\frac{e^{-r(T-t)}}{2\pi}\langle\hat{v}_{\rm min}|\varphi_T\rangle,
\label{eq:braket}
\ee
where 
\be 
|\varphi_T\rangle=\sum_{j_1,...,j_d=-N/2}^{N/2}\varphi_T(-\eta \vec{j}-i\alpha)|\vec{j}\rangle,
\ee
and
\be
|\hat{v}_{\rm min}\rangle=\sum_{j_1,...,j_d=-N/2}^{N/2}\hat{v}_{\rm min}(-\eta \vec{j}-i\alpha)|\vec{j}\rangle.
\ee
For a single asset, working in Fourier space and shifting
the integral sufficiently far off the real axis yields tremendous
improvements in numerical accuracy\cite{lewis}. We next show how to extend the speed
and precision of the Fourier approach into the multi-asset regime. In
order to make use of the quantum-inspired approach, we need an efficient
representation of the vectors $|\varphi_T\rangle$ and $|\hat{v}_{\rm
min}\rangle$ and an efficient way to perform the inner product in
Eq.~(\ref{eq:braket}), as well as a way to efficiently prepare the
vectors.  If the vectors can be accurately represented as nonnormalized
matrix product states (also known as tensor trains \cite{TT}), then the former
will be satisfied. However, the latter is more challenging. Here, we
adopt the black-box TT-cross algorithm \cite{TTcross1,TTcross2,TTcross3}
for efficiently preparing the states. Another candidate black-box
solution to the problem is ``tensor completion''
\cite{tensorcompletion}. 
For specific models, there are more reliable and faster solutions, for
instance by solving the Fokker Planck equation using density matrix
renormalization group (DMRG) methods, which we plan to explore in more
detail elsewhere.

\subsection{Matrix Product States}

A matrix product state (MPS) is a parametrized class of states (vectors
normalized in $\ell_2$ norm) on an $N^d$ dimensional Hilbert space. These
states are completely described by a set of $D\times D$ matrices
$\{A^{j_\alpha}\}$, with $j=1,...,N$ and $\alpha=1,...,d$ as:
\be 
|\Psi\rangle = \sum_{j_1,...,j_d} A^{j_1}\cdots A^{j_d}|j_1,\cdots,j_d\rangle.
\ee
The MPS vector is therefore described completely by $2NdD^2$ real
numbers, instead of $2N^d$ for the dense vector. Furthermore, the inner
product between two MPS with identical bond dimension can be evaluated
in ${\mathcal O}(2NdD^2)$ operations. In our case, each site represents
an asset, so $j_\alpha=-N/2,\cdots,N/2$. The matrices $\{A^{j_\alpha}\}$
carry the correlation between the assets for $|\varphi_T\rangle$. In
numerical simulations, the stronger the correlations, the larger the
{\em bond dimension} $D$ must be. In principle, in order to describe a
random quantum state in a $d$-dimensional Hilbert space exactly, a bond
dimension of $D=N^{d/2}$ is required, which is no simplification on the
dense encoding. However, in most practical settings, including the one
at hand, a small (constant) bond dimension suffices. In our numerical
experiments in Sec.~\ref{sec:bd}, we have observed that a bond dimension
of approximately $D=10$ --- $15$ is often sufficient to capture the
correlations of an arbitrary Gaussian state.

Finally, note that MPSs form the basis of the highly-effective DMRG
algorithm in quantum many body physics \cite{schollwock} and in quantum
chemistry \cite{garnet}. They have also been applied to various problems
in numerical and data analysis under the name of tensor trains
\cite{TT}.

\subsection{The TT-cross algorithm}

The TT-cross algorithm \cite{TTcross1} constructs an MPS representation
of a multi-variate function with relatively few calls to the function.
For example, a function $f:\mathbb{R}^d\rightarrow \mathbb{R}$ of $d$
variables can be discretized as a tensor with $d$ indices. If each
variable is discretized on $N$ grid points, the tensor takes $N^d$ real
values.  Clearly, without strong promises on sparsity or rank, even for
moderately large $N$ and $d$, the tensor cannot be written to memory or
manipulated efficiently. If it can be accurately approximated by an MPS
with small bond dimension, then the tensor becomes manageable.  In
general, it is not obvious how to build MPS representations just from
oracular access to the function $f$. One strategy is to build the MPS by
successive singular value decompositions and low-rank approximations of
the full tensor \cite{TTcross1}. The TT-cross algorithm solves this
problem in a different way.  The algorithm allows to approximate a
nonnormalized MPS of specified bond dimension $D$ by accessing the
function ${\mathcal O}(d N D^2)$ times \cite{TTcross2}, which can provide
tremendous time savings, when $d$ and $N$ are large. The main benefit of
the TT-cross algorithm over other black box constructions is that it
builds the tensor out of actual entries of the original function, making
it very well suited when oracular access to the function is
computationally cheap. When, instead of arbitrary oracular access we are
only given (unstructured) samples of the function, then tensor
completion \cite{tensorcompletion} is a better method of reconstructing
the MPS from the data.

The main idea is to perform an iterated matrix cross optimization. The
matrix cross identity states that a rank-$r$ $n\times m$ matrix $M$ can
be decomposed as \be M=M(:,J)M(I,J)^{-1}M(I,:),\ee where $J$, $I$ are
sets of $r$ columns and rows, $M(:,J)$ is the matrix of $r$ columns
indexed by $J$, and $M(I,J)$ is the sub-matrix of $M$ indexed by columns
$J$ and rows $I$. When $r\ll m,n$, the representation is more economical. The expression is exact for a rank $r$ matrix and
holds true for any choice of rows and columns.  It is natural to ask
what the best rank $r$ cross approximation is, when the matrix has rank
larger than $r$. The answer is known to be the submatrix with the
largest sub-volume $|\det(M(I,J))|$. Unfortunately, finding the largest
submatrix is an NP-hard problem, so we need to resort to approximations.

A good heuristic approach to finding the largest volume is to start with
a set of random columns $J$ , and obtain the $r$ rows leading to the
largest volume submatrix of $M(:,J)$. This can be done efficiently using
the maxvol algorithm \cite{maxvol}. Then starting from the rows found in
the previous step, find the new columns that optimize $M(I,:)$.
Performing this step iteratively quickly converges to a solution which
is often close to the optimum.

The extension to rank-$D$ tensors is straightforward. Consider an MPS
with $d$ sites, physical dimension $N$ and bond dimension $D$. Choose a
set of rows $\{I_1,...,I_d\}$ and columns $\{J_1,...,J_d\}$, where each
set $I_\alpha,J_\alpha$ has $D$ elements chosen from $N$. Then, sweeping
back and forth we can solve each matrix cross problem individually using
the above heuristic method by keeping all other links in the tensor
network fixed. In numerical experiments, it has been observed that the
process converges after a small number of sweeps.  Convergence is
measured by generating $50000$ random sample points of the function and
of the MPS approximation to the function, and evaluate the one-norm
difference between the two.  For a more comprehensive explanation of the
TT-cross algorithm with pseudo-code, see Refs.~\cite{TTcross2} and
\cite{TTcross3}.

\section{Numerical experiments} 
\label{sec:num}

We now show that the tensor network approach to multi-asset Fourier
options pricing faithfully extends the great precision of Fourier
options pricing into the multi-asset regime. In all of our experiments,
we set the parameters of the model to $r=0.3$, $\sigma=0.5$, and
$T=1$. The strike price and starting asset price are set to $S_0=K=100$
homogeneously across all assets. These are fairly standard settings and
serve to illustrate the general behaviour of the tensor network solver.
The conclusions do not change qualitatively with different choices of
parameters.  Simulations are performed in python on a standard laptop
without GPU acceleration. The TT-cross components used the package
\texttt{tntorch} \cite{tntorch}.

\subsection{The single asset case}

\begin{figure}[h]
\includegraphics[width=\linewidth]{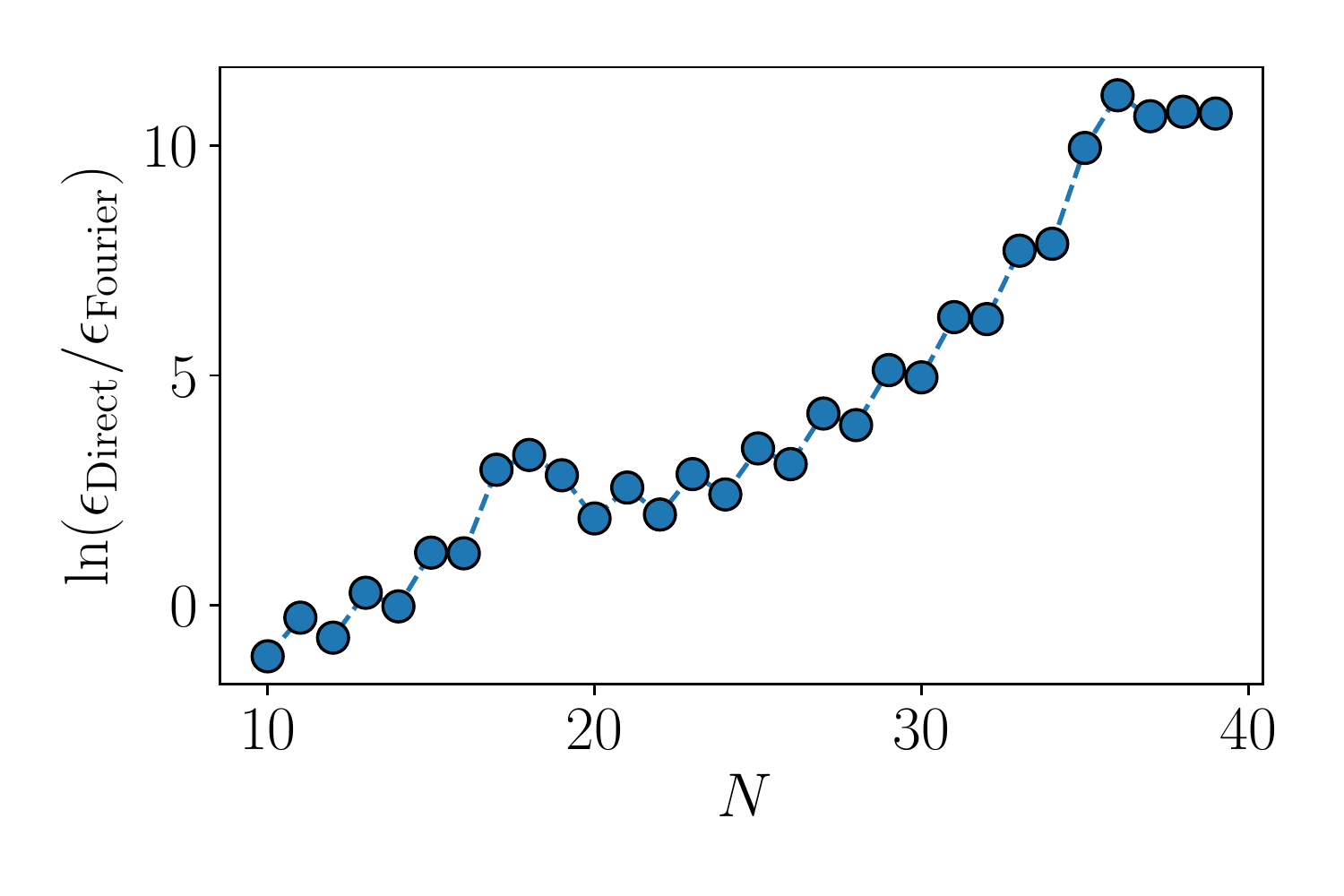}
\caption{Logarithm of the relative accuracy $\epsilon_{\rm
direct}/\epsilon_{\rm Fourier}$ of the discretized direct formula
[Eq.~(\ref{eq:mc})] with respect to the discretized Fourier formula
[Eq.~(\ref{eq:maincont})], for a fixed number of grid points $N$. Note
that the relative error increases exponentially with $N$.}
\label{fig1}
\end{figure}

We set the stage by reviewing and comparing the Fourier options pricing
method with vanilla Monte Carlo for the European call option under
geometric Brownian motion. These results are already well established
but serve as a baseline for extension to the multi-asset case.  In the
single asset case, we compare the convergence of the Monte Carlo
algorithm in the number of samples to the Fourier method
[Eq.~(\ref{eq:maincont})] and the \textit{direct} method
[Eq.~(\ref{eq:mc})] in the number of interpolation points.  Figure
\ref{fig1} shows the relative accuracy $\epsilon_{\rm
direct}/\epsilon_{\rm Fourier}$ with
$\epsilon_{\kappa}=\frac{|V_\kappa-V_{\rm exact}|}{V_{\rm exact}}$, and
$\kappa=\{ {\rm direct},{\rm Fourier}\}$.  For the same number of grid
points, the Fourier method dramatically outperforms the direct
integration.  While the direct integration converges with an inverse
polynomial, the Fourier method converges faster than polynomially and
likely exponentially in the number of integration points.  For the
Fourier options simulation, we use $\eta = 0.5$ and $\alpha=3$ as fixed
hyperparameters. The two kinks in the data are artefacts of the  method.
They depend on the choice of $\eta$. If $\eta$ is too large, the kinks
become more pronounced, but the exponential scaling in $N$ levels off
before reaching computer precision limits. The parity of $N$ also plays
a role in the accuracy of the Fourier method. For further analysis of
the Fourier method on a single asset, see Ref.~\cite{schmelzle} and
references therein.

In this experiment, the vanilla Monte Carlo approach reaches a standard
deviation of $10^{-4}$ for $10^{10}$ --- $10^{11}$ samples, which is
impractical on a laptop. For comparison, the direct integration requires
$N\geq 100$ grid points, and the Fourier method requires only $N\geq30$
grid points, both running in milliseconds on a laptop. Therefore, the
Fourier methods provides remarkable speedup over the other two methods
in the single asset case. We now demonstrate that this advantage can be
extended to the multi-asset case by using the tensor network approach
outlined above.

\subsection{The multi-asset case}

In the multi-asset case, we consider the minimum payoff
[Eq.~(\ref{eq:minpayoff})] and a parametrized family of correlated
geometric Brownian motion with correlation matrix:
\be 
\Sigma(\beta) = (\beta |+\rangle\langle +| + \1)/(1+\beta),
\label{eq:covmat}
\ee
where we restrict $0\leq \beta\leq 1$. Here, the state
$|+\rangle=\sum_{j=1}^d |j\rangle$, and the correlation matrix has
$\beta/(1+\beta)$ on the off diagonals, and ones on the diagonal. This
choice of covariance matrix, although artificial, captures with a single
parameter the amount of correlation in the dynamics of the problem. Our
general conclusions extend to other correlated dynamics. We set the
number of grid points to $N=50$, which guarantees an accuracy well
within $10^{-4}$. The imaginary shift is chosen to be $\alpha=5/d$,
which satisfies the requirements ${\rm Im}[z_j]>0$ and $\sum_{j=1}^d
{\rm Im}[z_j]>1$, and allows for good stability of the TT-cross
algorithm while guaranteeing rapid convergence.

\begin{table}
\caption{Runtime and accuracy results for a single typical run of the
tensor-Fourier algorithm on the min-option for multi-asset geometric
Brownian motion with a covariance matrix given in Eq.~(\ref{eq:covmat})
with $\beta=0.5$. $t_{\rm wall}$ is the wall clock time, $t_{\rm rel}$
is the ratio of the wall clock time for the tensor-Fourier method over
vanilla Monte Carlo, $r_{\rm comp}$ is the compression ratio of the
computation, $\epsilon_{\rm trunc}$ is the MPS truncation error compared
with the full tensor, $D_w$ and $D_\varphi$ are the bond dimensions of
the two MPS, and $\eta$ is the grid spacing in momentum space. }
\label{tab:results}
\begin{tabular}{ l l c c c c c c r } 
\hline\hline 
  $d$ & $t_{\rm wall}$[s] & $t_{\rm rel}$ & $r_{\rm comp}$ & $\epsilon_{\rm trunc}$ & $D_v$ & $D_\varphi$ & $\eta$ \\ 
 \hline
 $2$ & $0.027$ & $3\times 10^{-5}$ & $3.6$&$1.42 \times 10^{-6}$ & $20$ & $10$ & $0.5$ \\ 
 $3$ & $0.73$ & $5.4\times 10^{-4}$ &$1.056$& $4.10 \times 10^{-6}$ & $20$ & $10$ & $0.4$ \\ 
 $4$ & $6.8$ & $0.0038$ &$0.074$& $1.84 \times 10^{-6}$ & $30$ & $15$ & $0.3$ \\ 
 $5$ & $10.2$ & $0.0045$ &$0.0022$&---& $30$ & $15$ & $0.3$ \\ 
 $6$ & $13.7$ & $0.0051$ &$5.85 \times 10^{-5}$&---& $30$ & $15$ & $0.2$ \\ 
 $7$ & $52.8$ & $0.017$ &$2.58 \times 10^{-6}$&---& $40$ & $20$ & $0.2$ \\ 
$8$ & $63.4$ & $0.018$ &$6.19 \times 10^{-8}$&---& $40$ & $20$ & $0.2$ \\ 
$9$ & $74.0$ & $0.018$ &$1.44 \times 10^{-9}$&---& $40$ & $20$ & $0.2$ \\ 
$10$ & $84.5$ & $0.019$ & $3.30 \times 10^{-11}$&---& $40$ & $20$ & $0.2$ \\ 
$15$ & $326.8$ & $0.048$ &$2.67 \times 10^{-19}$&---& $50$ & $25$ & $0.2$ \\ 
\hline\hline
\end{tabular}
\end{table}

We simulate the correlated multi-asset option with the tensor Fourier
algorithm based on the TT-cross algorithm described in
Sec.~\ref{sec:TN}, the correlated multi-asset option with the full
summation of Eq.~(\ref{eq:main}) up to $d=4$, and the correlated
multi-asset option with a vanilla Monte Carlo simulation
(Sec.~\ref{sec:MC}). A typical single shot run is reported in Table
\ref{tab:results}. $D_v$ is the bond dimension used in TT-cross
algorithm for the MPS approximation of $|\hat{v}\rangle$, and
$D_\varphi$ is the bond dimension used in the TT-cross algorithm for the
MPS approximation of $|\varphi_T\rangle$. Both are chosen so as to
guarantee convergence of TT-cross algorithm in a single sweep with
convergence tolerance $\epsilon_{\rm tol}=0.005$. We find empirically
that allowing for several sweeps does not improve the convergence
significantly, nor does progressively increasing the bond dimension at
each sweep. $t_{\rm wall}$ is the wall clock time in seconds for the
tensor-Fourier algorithm. $r_{\rm comp}$  is the compression ratio,
defined as the number of oracular function calls made by the TT-cross
algorithm divided by the total grid size $N^d$. The truncation error
$\epsilon_{\rm trunc}$ is defined as the ratio between the options price
calculated in the MPS approximation generated via the TT-cross algorithm
and the full tensor.  The comparison already becomes infeasible for
$d=5$ without further compression tricks. The relative time $t_{\rm
rel}$ is defined as the ratio of wall clock time for the tensor Fourier
method over the wall clock time for vanilla Monte Carlo with $n=5\times
10^7$ samples, which guarantees precision up to $10^{-3}$.  With $50$
grid points and the hyperparameters indicated in the table, we expect
the tensor-Fourier results to reach a precision well within $10^{-4}$.
Hence, the  $t_{\rm rel}$ speedup is a conservative estimate. The wall
clock time for Monte Carlo is calculated on the same hardware as the
tensor-Fourier simulations for $d=2$. For $d>2$, the Monte Carlo wall
clock time is extrapolated assuming a  linear scaling with the dimension
$d$. This estimate is also conservative, because the Monte Carlo
simulations scale slightly super-linearly. The comparison with Monte
Carlo is meant to give an indication of the ballpark computational time for
the tensor-Fourier method. A much more detailed analysis would be
required to compare an optimal Monte Carlo method, such as multilevel
Monte Carlo \cite{mlmc}, to an optimal tensor-Fourier method, using
perhaps the COS transform \cite{COS}. Furthermore, the relative
advantages of different methods depends strongly on the application at
hand. Our interest is in showing the feasibility and potential
advantages of the tensor-Fourier method.

The wall clock time $t_{\rm wall}$ depends sensitively on $D_v$ and
$D_\varphi$, and only weakly on $d$. The runtime of the TT-cross
algorithm is expected to scale as ${\mathcal O}(dND^2)$ \cite{TTcross1}.
Still, the runtime is modest up to large dimensionality, maintaining
several order of magnitude speedup over the vanilla Monte Carlo
approach. The compression ratio $r_{\rm trunc}$ gives an idea of how
efficiently the MPS can be constructed via the TT-cross algorithm. The
number of oracular calls to the integrand scales as ${\mathcal
O}(ND^2)$. The truncation error $\epsilon_{\rm trunc}$ does not appear
to depend sensitively on $d$. Rather, it is determined by the choice of
convergence criterion of the TT-cross algorithm.

Ultimately, the strength of the Fourier approaches is the ability to
reach extremely high precision in a reasonable time. Our numerical
results indicate that this is the case up to at least $d=15$ assets. We
find that beyond $d=15$, the TT-cross algorithm has trouble converging.
This is likely caused by a proliferation of local minima and rugged
optimization landscapes. It is an interesting open research question,
whether there are good schemes for providing the TT-cross algorithm with
a warm start for larger systems or to embed it into a meta-heuristic to
overcome the ruggedness of the landscape.

\subsection{The bond dimension}
\label{sec:bd}

\begin{figure}
\includegraphics[width=1.0\linewidth]{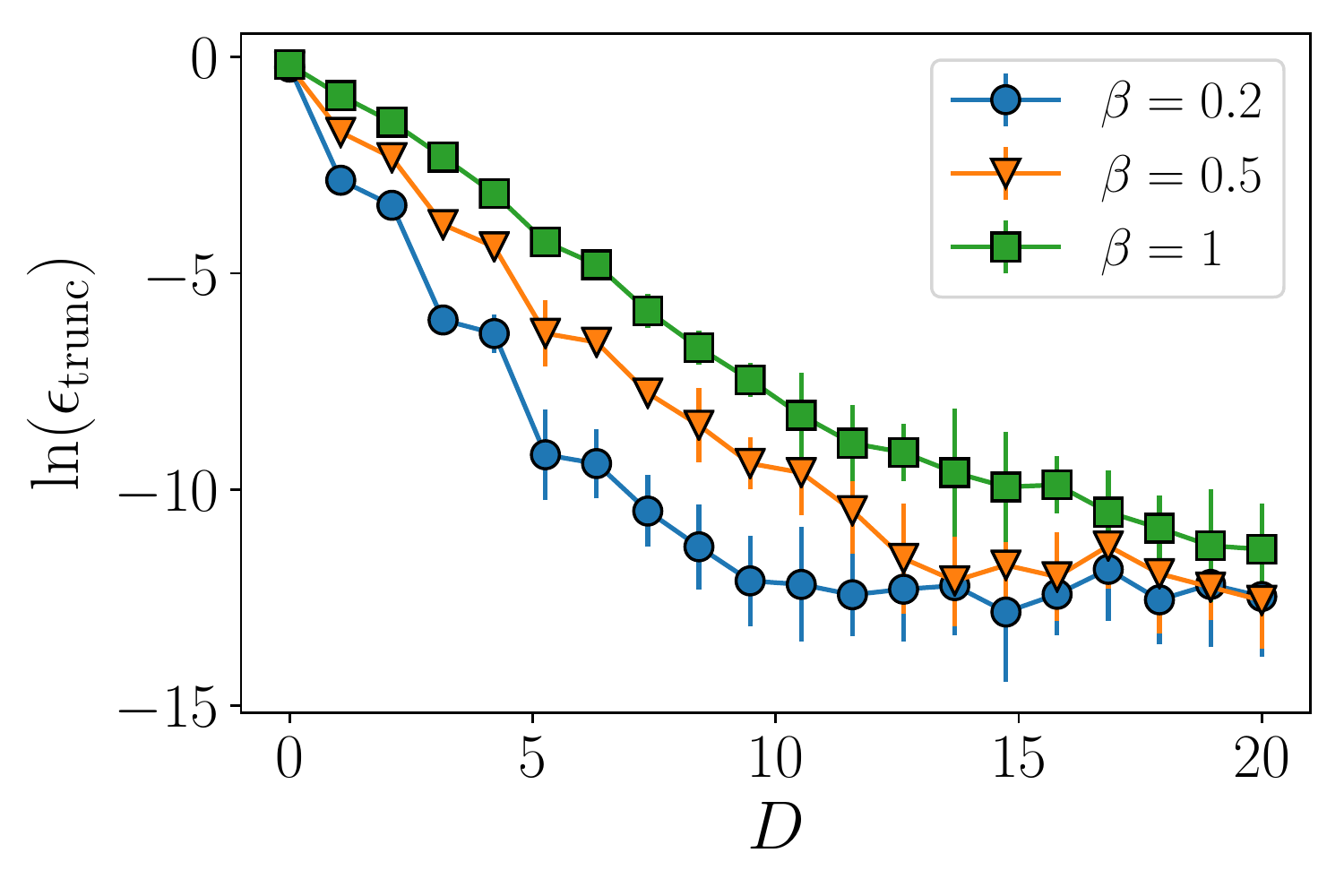}
\caption{Logarithm of the truncation error as a function of the bond 
dimension $D_\varphi$ for $\beta=0.2$, $0.5$,$ 1$, and $d=3$.
\label{fig:2}}
\end{figure}

The wall clock time, the compression ratio and the truncation error all
depend sensitively on the choice of bond dimensions. We discuss how how
the bond dimension of the MPS for the characteristic function
$|\varphi_T\rangle$ and the payoff function $|\hat{v}_{\rm min}\rangle$
depends on the various hyperparameters of the system.

The tensor network solver relies heavily on the TT-cross subroutine
which is heuristic and involves stochastic components. In particular,
the random staring point for the TT-cross algorithm affects the
convergence significantly. What we observe in our experiments is that,
when the bond dimension is sufficiently large (although not too large),
the TT-cross algorithm reliably converges. When the bond dimension is
taken too small, TT-cross algorithm does not converge. In the
intermediate regime, the starting point is essential to guarantee
convergence. However, we notice that the quality of the approximation
can be reasonably good even though TT-cross algorithm does not converge.
This is not inconsistent, because the convergence criterion reflects the
actual proximity of the original vectors and their MPS approximation,
while we are interested in functionals of the vectors.

This can be seen in Fig.~\ref{fig:2} where the convergence to
non-convergence transition happens for $D\approx 10 \beta$ for a
convergence criterion of $\epsilon=0.005$, but the error in the
approximation of the function decreases exponentially in $D$. This
suggests that our algorithm smoothly improves with $D$. Figure
\ref{fig:2} also shows the logarithm of the truncation error for a $d=3$
experiment for different intensities of correlations. The reported
results are averaged over $20$ runs. The bond dimension for
$|\hat{v}_{\rm min}\rangle$ is taken to be $D_v=30$ to guarantee an
accurate representation of the tensor. Hence all of the inaccuracy is
caused by the MPS truncation error of the characteristic function
$|\varphi_T\rangle$.

We consistently observe that the accuracy improves exponentially with
the bond dimension $D_\varphi$, with an exponent inversely proportional
to $\beta$. The precision saturates at a level close to
$\log(\epsilon_{\rm trunc})=-12$. We do not know what causes the
saturation, but we expect that is has to do with TT-cross algorithm
reaching local minima when the bond dimension is taken to be large.

\section{Discussion} 

We have shown how to use tensor network methods to extend the
applicability of Fourier options pricing to the many asset setting. This
promises to dramatically reduce the computational time of multi-asset
options pricing, in a similar way as is already the case for single
asset options. The natural question remains as to when the tensor
network approach is applicable. There are two natural extensions to be
considered for practical applications. First, to adapt the method to
more complex multi-asset instruments and second, to adapt it to
different forms of stochastic dynamics.

The forms of stochastic dynamics that naturally fit into our framework
are those for which the (multivariate) characteristic function is known
analytically. This includes, among others, the Merten jump model
\cite{lewis} and the Heston stochastic volatility model \cite{heston};
two of the most commonly used models in options pricing.  Other models
such as the SABR model \cite{SABR} do not have a know analytic form of
the characteristic function, but there exist methods to extend Fourier
options pricing to them. More complex multi-asset instruments could be
challenging to generalise, yet we already know how to extend some common
ones such as basket and rainbow options \cite{basket}. Time-dependent
options can also be treated following ideas taken from
Ref.~\cite{path_dep_op}. Therefore, the tensor Fourier approach holds
the promise of accelerating many common multi-asset derivatives pricing
models. How great the speedup is will ultimately depend on the nature
and complexity of the given instrument.

\begin{acknowledgments}

We especially thank Helmut G.~Katzgraber for thoroughly reviewing the manuscript and providing helpful recommendations. We thank Martin Schuetz, Grant Salton and Henry
Montague for helpful discussions.

\end{acknowledgments}

\bibliographystyle{unsrt}
\bibliography{biblio}

\begin{thebibliography}{10}

\bibitem{hullBook}
John~C Hull.
\newblock {\em Options futures and other derivatives}.
\newblock Pearson Prentice Hall, Upper Saddle River, NJ, 2006.

\bibitem{quasiMC}
Peter~A Acworth, Mark Broadie, and Paul Glasserman.
\newblock A comparison of some monte carlo and quasi monte carlo techniques for
  option pricing.
\newblock In {\em Monte Carlo and Quasi-Monte Carlo Methods 1996}, pages 1--18.
  Springer, 1998.

\bibitem{mlmc}
Michael~B Giles.
\newblock Multilevel monte carlo methods.
\newblock {\em Acta Numerica}, 24:259--328, 2015.

\bibitem{benchop}
Lina von Sydow, Lars Josef~H{\"o}{\"o}k, Elisabeth Larsson, Erik Lindstr{\"o}m,
  Slobodan Milovanovi{\'c}, Jonas Persson, Victor Shcherbakov, Yuri
  Shpolyanskiy, Samuel Sir{\'e}n, Jari Toivanen, et~al.
\newblock Benchop--the benchmarking project in option pricing.
\newblock {\em International Journal of Computer Mathematics},
  92(12):2361--2379, 2015.

\bibitem{Q1}
Nikitas Stamatopoulos, Daniel~J Egger, Yue Sun, Christa Zoufal, Raban Iten,
  Ning Shen, and Stefan Woerner.
\newblock Option pricing using quantum computers.
\newblock {\em Quantum}, 4:291, 2020.

\bibitem{Q2}
Patrick Rebentrost, Brajesh Gupt, and Thomas~R Bromley.
\newblock Quantum computational finance: Monte carlo pricing of financial
  derivatives.
\newblock {\em Physical Review A}, 98(2):022321, 2018.

\bibitem{Q3}
Shouvanik Chakrabarti, Rajiv Krishnakumar, Guglielmo Mazzola, Nikitas
  Stamatopoulos, Stefan Woerner, and William~J Zeng.
\newblock A threshold for quantum advantage in derivative pricing.
\newblock {\em Quantum}, 5:463, 2021.

\bibitem{Q4}
Javier Gonzalez-Conde, {\'A}ngel Rodr{\'\i}guez-Rozas, Enrique Solano, and
  Mikel Sanz.
\newblock Pricing financial derivatives with exponential quantum speedup.
\newblock {\em arXiv preprint arXiv:2101.04023}, 2021.

\bibitem{Q5}
Koichi Miyamoto.
\newblock Bermudan option pricing by quantum amplitude estimation and chebyshev
  interpolation.
\newblock {\em arXiv preprint arXiv:2108.09014}, 2021.

\bibitem{montanaro}
Ashley Montanaro.
\newblock Quantum speedup of monte carlo methods.
\newblock {\em Proceedings of the Royal Society A: Mathematical, Physical and
  Engineering Sciences}, 471(2181):20150301, 2015.

\bibitem{childs}
Andrew~M Childs, Jin-Peng Liu, and Aaron Ostrander.
\newblock High-precision quantum algorithms for partial differential equations.
\newblock {\em Quantum}, 5:574, 2021.

\bibitem{montanaroFEM}
Ashley Montanaro and Sam Pallister.
\newblock Quantum algorithms and the finite element method.
\newblock {\em Physical Review A}, 93(3):032324, 2016.

\bibitem{TTcross1}
Ivan Oseledets and Eugene Tyrtyshnikov.
\newblock Tt-cross approximation for multidimensional arrays.
\newblock {\em Linear Algebra and its Applications}, 432(1):70--88, 2010.

\bibitem{blackscholes}
Fischer Black and Myron Scholes.
\newblock The pricing of options and corporate liabilities.
\newblock {\em The Journal of Political Economy}, 81(3):637--654, 1973.

\bibitem{SABR}
Patrick~S Hagan, Deep Kumar, Andrew~S Lesniewski, and Diana~E Woodward.
\newblock Managing smile risk.
\newblock {\em The Best of Wilmott}, 1:249--296, 2002.

\bibitem{barrier}
H{\'e}lyette Geman and Marc Yor.
\newblock Pricing and hedging double-barrier options: A probabilistic approach.
\newblock {\em Mathematical finance}, 6(4):365--378, 1996.

\bibitem{lewis}
Alan~L Lewis.
\newblock A simple option formula for general jump-diffusion and other
  exponential l{\'e}vy processes.
\newblock {\em Available at SSRN 282110}, 2001.

\bibitem{heston}
Steven~L Heston.
\newblock A closed-form solution for options with stochastic volatility with
  applications to bond and currency options.
\newblock {\em The review of financial studies}, 6(2):327--343, 1993.

\bibitem{van2018cos}
Zaza van~der Have and Cornelis~W Oosterlee.
\newblock The cos method for option valuation under the sabr dynamics.
\newblock {\em International Journal of Computer Mathematics}, 95(2):444--464,
  2018.

\bibitem{carrmadan}
Peter Carr and Dilip Madan.
\newblock Towards a theory of volatility trading.
\newblock {\em Option Pricing, Interest Rates and Risk Management, Handbooks in
  Mathematical Finance}, pages 458--476, 2001.

\bibitem{schmelzle}
Martin Schmelzle.
\newblock Option pricing formulae using fourier transform: Theory and
  application.
\newblock {\em Preprint, http://pfadintegral. com}, 2010.

\bibitem{eberlein}
Ernst Eberlein, Kathrin Glau, and Antonis Papapantoleon.
\newblock Analysis of fourier transform valuation formulas and applications.
\newblock {\em Applied Mathematical Finance}, 17(3):211--240, 2010.

\bibitem{TT}
Ivan~V Oseledets.
\newblock Tensor-train decomposition.
\newblock {\em SIAM Journal on Scientific Computing}, 33(5):2295--2317, 2011.

\bibitem{TTcross2}
Lev~I Vysotsky, Alexander~V Smirnov, and Eugene~E Tyrtyshnikov.
\newblock Tensor-train numerical integration of multivariate functions with
  singularities.
\newblock {\em arXiv preprint arXiv:2103.12129}, 2021.

\bibitem{TTcross3}
Sergey Dolgov and Dmitry Savostyanov.
\newblock Parallel cross interpolation for high-precision calculation of
  high-dimensional integrals.
\newblock {\em Computer Physics Communications}, 246:106869, 2020.

\bibitem{tensorcompletion}
Qingquan Song, Hancheng Ge, James Caverlee, and Xia Hu.
\newblock Tensor completion algorithms in big data analytics.
\newblock {\em ACM Transactions on Knowledge Discovery from Data (TKDD)},
  13(1):1--48, 2019.

\bibitem{schollwock}
Ulrich Schollw{\"o}ck.
\newblock The density-matrix renormalization group in the age of matrix product
  states.
\newblock {\em Annals of physics}, 326(1):96--192, 2011.

\bibitem{garnet}
Garnet Kin-Lic Chan and Sandeep Sharma.
\newblock The density matrix renormalization group in quantum chemistry.
\newblock {\em Annual review of physical chemistry}, 62:465--481, 2011.

\bibitem{maxvol}
Sergei~A Goreinov and Eugene~E Tyrtyshnikov.
\newblock The maximal-volume concept in approximation by low-rank matrices.
\newblock {\em Contemporary Mathematics}, 280:47--52, 2001.

\bibitem{tntorch}
https://github.com/vmml/tntorch.

\bibitem{COS}
Fang Fang and Cornelis~W Oosterlee.
\newblock A novel pricing method for european options based on fourier-cosine
  series expansions.
\newblock {\em SIAM Journal on Scientific Computing}, 31(2):826--848, 2009.

\bibitem{basket}
Ruggero Caldana, Gianluca Fusai, Alessandro Gnoatto, and Martino Grasselli.
\newblock General closed-form basket option pricing bounds.
\newblock {\em Quantitative Finance}, 16(4):535--554, 2016.

\bibitem{path_dep_op}
Fang Fang and Cornelis~W Oosterlee.
\newblock A fourier-based valuation method for bermudan and barrier options
  under heston's model.
\newblock {\em SIAM Journal on Financial Mathematics}, 2(1):439--463, 2011.

\end{thebibliography}

\end{document}